# A Modular High Temperature Measurement Set-Up for Semiconductor Device Characterization

Peter Borthen and Gerhard Wachutka
Institute for Physics of Electrotechnology, Munich University of Technology,
80290 München, Germany

*Abstract*—We demonstrate the capabilities of a high temperature measurement set-up recently developed at our institute. It is dedicated to the characterization of semiconductor devices and test structures in the temperature range from room temperature up to 500°C and higher. A detailed description of the experimental equipment is given. Its practical use is demonstrated by measuring temperature-dependent characteristics of silicon VDMOSFET and IGBT devices as well as SiC-diodes. For the silicon devices, numerical simulations based on recently developed high temperature physical models were also performed in order to gain a deeper understanding of the measured data, together with a revalidation of the model parameters.

## I. Introduction

Electronic devices and modules are increasingly employed in very hot environments such as combustion engines, car brake systems, turbines or well drilling. Therefore, the high temperature characterization of semiconductor devices is getting more and more practical relevance. However, since commercial probe stations do usually not allow for measurements at very high temperatures, semiconductor devices developed to operate in this temperature range cannot be properly characterized using standard equipment. Up to now, only little experimental equipment suited for the electrothermal analysis of semiconductor devices at very high temperatures has been reported. Reggiani at al. [1,2] used a modified commercial high temperature oven for measurements on samples at temperatures as high as 800°C. RF measurements on wafers at temperatures up to 500°C were described in [3].

With continuously decreasing size and increasing complexity of semiconductor devices, device simulations play an essential role for their optimal design. Device simulations strongly rely on the quality of the physical models implemented in the applied software. This is particularly true for the simulation of high temperature device properties. Till lately, there were no validated models available for temperatures above 250°C [4]. Only in recent work the temperature range for calibrated models has been extended to about 400°C for the mobility models and even 800°C for the impact ionization coefficients [1,2,5]. We used these new models for the simulation of the high temperature device behavior described below.

In the following we describe a versatile, dedicated experimental set-up for the electrical and thermal characterization of semiconductor devices and test structures. The measurements are performed under high vacuum conditions which provides an excellent thermal isolation of the device under test and, at the same time, prevents the degradation of then sample and the system parts due to the exposure to the oxygen in the surrounding air. Apart from measuring device terminal characteristics, the extraction of material properties like carrier mobilities, generation-recombination rates, impact ionization rates, thermal conductivity and others may be performed. The results will be used in the refinement and calibration of physical device models required for the simulation of the high temperature device behavior.

## II. Experimental set-up

A schematic view of the experimental set-up is shown in Fig. 1. It consists of two main building blocks: A vacuum system and a set of measuring instruments controlled by a computer. The vacuum system comprises Trinos modular vacuum chambers [6] with an inner diameter of 40 cm and a Leybold PT 151 pumping unit [7] consisting of a rotary pump and a turbomolecular pump (Fig. 2). The lower one of the two chambers has several ports for electrical feedthroughs as well as a feedthrough for the cooling liquid.

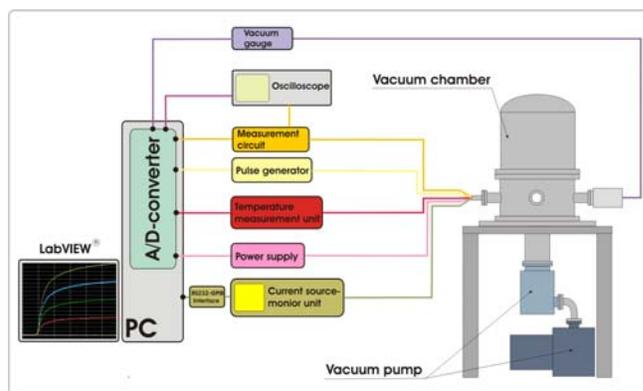

Figure 1. Block diagram of the high-temperature measurement set-up.

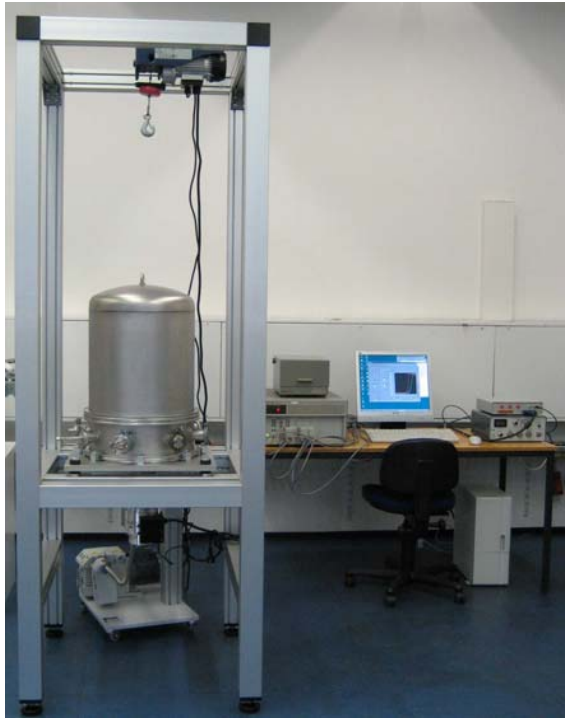

Figure 2. View of the vacuum system.

The chamber bell can be lifted by means of a motorized winch, thus facilitating the access to the measuring platform inside the vacuum chamber. The minimum achievable pressure in the vacuum chamber is about 1 mPa which is limited by the kind of the flanges used in our system (KF-flanges) as well as the size of the vacuum chambers and the pumps. However, this pressure is fully sufficient for the purpose of this system. The pumping station can alternatively be connected to the vacuum chamber of a second module for Hall measurement.

Inside the vacuum chamber, different exchangeable measuring platforms can be used depending on the kind of the measurement task. Currently, a general-purpose prober unit (Fig. 3) and a platform with a circuit for high current pulsed measurements (Fig. 4 and 5) are in use. Both systems are in-house developments.

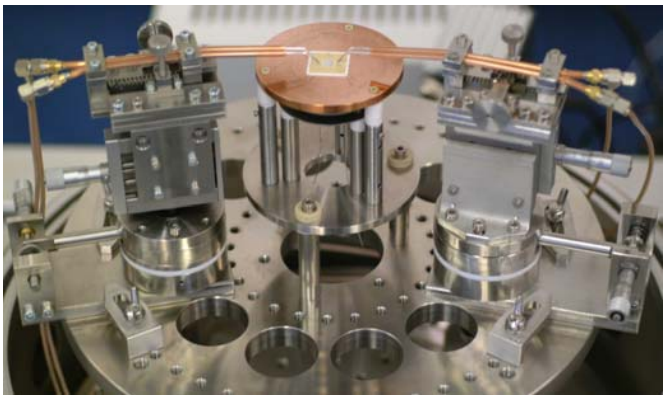

Figure 3. High-temperature sample prober

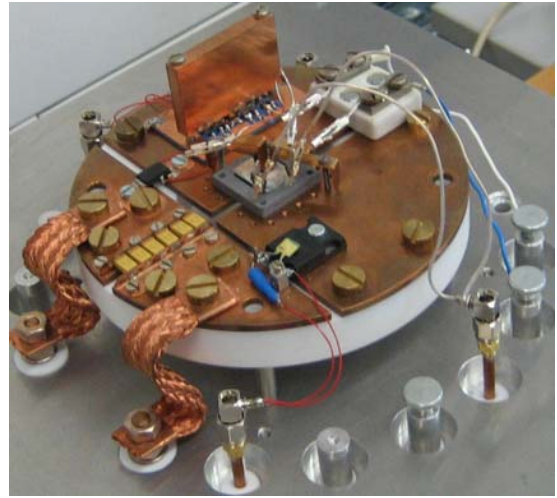

Figure 4. High-temperature circuit for pulsed measurements

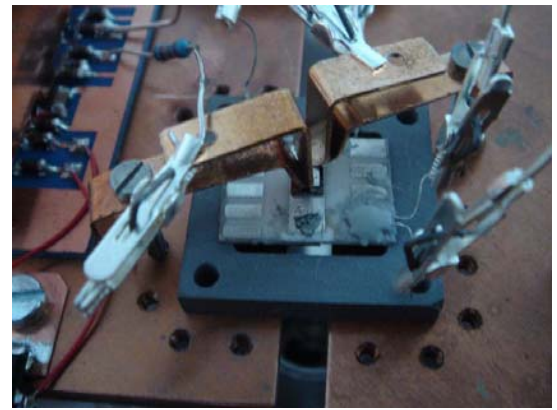

Figure 5. Detailed view of the sample ceramic substrate with electrical contacts. The heating coil is located below the substrate.

The sample stage of the general-purpose prober unit is made of a circular copper plate with a resistive heating coil underneath. With this unit, a maximum temperature of about 600 °C can be reached. For temperature measurement a PT100 platinum resistor gauge is used which is placed in the vicinity of the sample. Stainless steel manipulators with coax-cable (RG402) cantilevers fixed on top of the manipulators are used for contacting the samples. On one end of the cable (about 6 mm) the shielding and the PTFE-dielectric has been removed and the core (silver-plated steel, diameter 0.9 mm) has been sharpened to form a probe tip with a radius of approx. 100 µm. A temperature control and power supply circuit reads and converts the PT100 resistance values. All data is collected by a PC using a PCI-6014 A/D-converter (National Instruments). The whole heating system is controlled by a LabVIEW® program developed for the specific measurement requirements. Currently, another platform is being developed, on which by the use of an electron beam heater measurements up to temperatures of 800-1000°C will be achieved.

The electrical characterization of the samples is performed using an Agilent E5270B modular parameter analyzer. It is connected to the PC via an RS232/GPB-interface and controlled by LabVIEW® software.

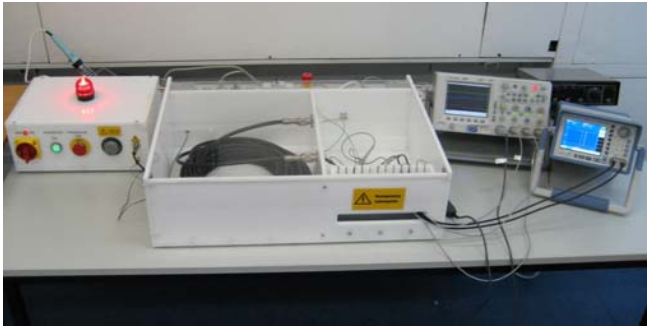

Figure 6. Experimental TLP-System

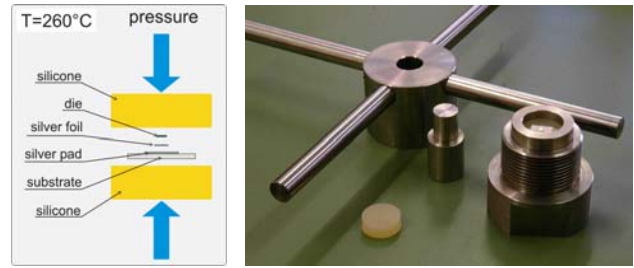

Figure 7. Principle of the NTV die bonding technique (left) and a simple manual press developed in-house (right), which allows for die bonding on substrates as large as 22x22 mm$^2$.

A number of different measurement modes such as staircase sweep or pulsed measurement can be selected for current-voltage characteristics. The E5270B parameter analyzer is very well suited for measurements in the low and very low current and voltage range. The maximum current in our present instrumental configuration is limited to 1.5 A and the shortest pulse length possible is 500µs.

With a view to extending the capabilities of our set-up to transient recording and higher terminal currents, as it is required for the analysis of power devices, a dedicated electronic circuitry had to be developed. Currently this circuit is able to generate current pulses of max. 200A and pulse lengths between 2 µs and 100 µs. The resistive heating coil is attached to the backside of the ceramic substrate holding the device under test by means of a two-component ceramic adhesive (Fig. 4 and 5).

For high-temperature analysis with very short pulses in the range of 10-500ns and voltage amplitudes up to several kilovolts a TLP-System is currently under development. Figure 6 shows an experimental version of the TLP-System used for exploratory tests [8]. This system will enable the investigation of breakdown phenomena in electronic devices under extreme temperature conditions.

### III. DIE-BONDING

The stability and robustness of the die attachment and the wire bonds is the crucial point for measurements at very high temperature. In order to achieve stable measurement conditions even for temperatures above 700 K, a special die bonding technique (described as NTV process in [9]) is employed. With this process, a very reliable bonding is achieved by applying high mechanical pressure on a porous silver foil placed between the die backside and the metallized substrate at elevated temperature (Fig. 7). The result is a thin sintered layer between die and substrate with excellent bonding properties over a wide range of temperature. Due to the very good electrical and thermal conductivity of silver, its high melting temperature and a well-adapted coefficient of thermal expansion, this bonding method seems to be the best choice for high-temperature test samples.

A relatively high pressure of about $4x10^7$ Pa and heating above 200°C was applied in the original work [9]. As a suitable press for such high mechanical pressure was not available to us, we developed a simple manual press which can be inserted into an oven for the heating step.

The maximum pressure achieved with this press was estimated to be about $1x10^7$ Pa. As this value is lower compared to that in [9], we had to raise the sintering temperature to about 260°C and extend the pressing time to about 1h in order to achieve comparable results. All samples bonded by this method proved to be mechanically and electrically very stable: There were no signs of failure even after repeated heating up to 500°C.

### IV. HIGH-TEMPERATURE MEASUREMENT AND SIMULATION ON TEST DEVICES

Figure 8 shows the measured and the simulated transfer characteristics, respectively, of a 75V MOSFET placed within a PCM test structure, which had been fabricated by means of the so-called SFET2 process technology available at Infineon. The PCM test structure comprises 38 MOSFET cells only, whereas the full device consists of about 500,000 cells.

As we can recognize from Fig. 8, the basic MOSFET operation stays stable (with negative temperature coefficient) up to a temperature of about 300°C. Above this temperature level, the drain current increases with temperature so that the device is likely to run into an electrothermal instability leading eventually to device failure. However, already between 200°C and 300°C we find a significant increase in the subthreshold current and a decrease of the threshold voltage, which is considerably larger than that at lower temperatures. This device degradation is mainly caused by the decreasing electron mobility and the increasing thermal generation of electron-hole pairs observed at rising lattice temperature.

It turns out that the high-temperature simulation models, which should be able to cover the full temperature range of the measured data, still need a recalibration of the model parameters. With the given temperature dependence in the physical models as implemented in DESSIS, it is virtually impossible to find a single parameter set that would perfectly describe the device operation for all temperature values between room temperature and 400°C. An acceptable agreement between measurement and simulation could only be achieved after re-adjusting the temperature dependence in a number of model parameters. Actually, this re-adjustment had to be made for the parameters of band gap narrowing, channel mobilities, and generation-recombination processes.

The next device type investigated in this work was a 600 V NPT (*non-punch-through*) IGBT. Due to their specific structure, IGBTs show a high-temperature behaviour that combines that of a MOSFET with that of a BJT. As a result, the forward characteristics exhibit two voltage regions with a positive, BJT-type, and a negative, MOSFET-type temperature coefficient. There exists also one operating point ($U_{CE}$, $I_{CE}$) in the characteristics which is nearly independent of temperature. Measured and simulated transfer characteristics are shown in Figure 10.

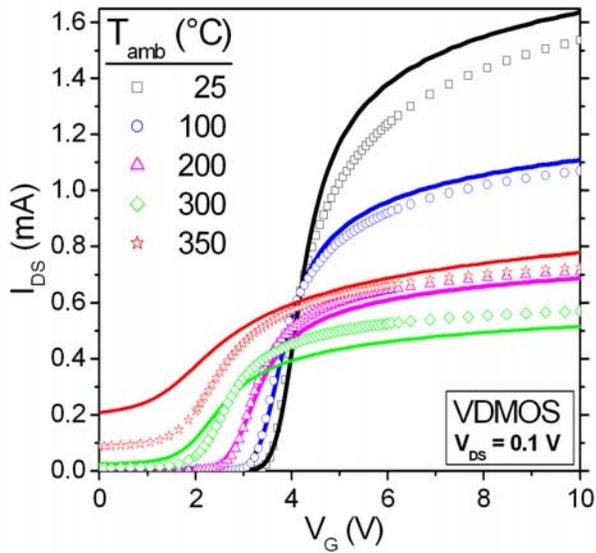

Figure 8. Transfer characteristics of a silicon VDMOSFET test structure. Solid lines: Measurement; Symbols: Simulation.

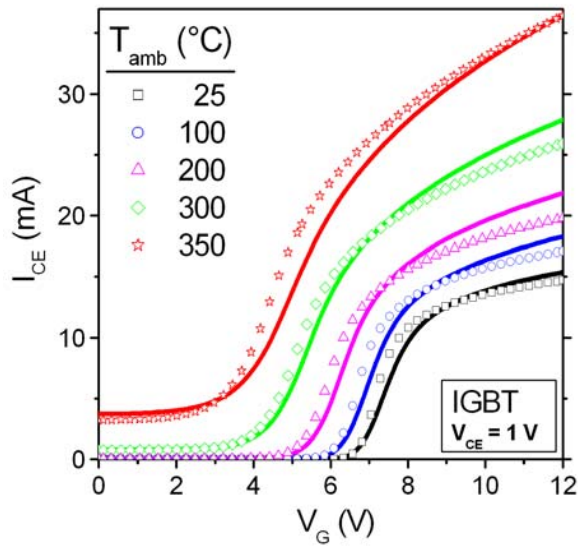

Figure 9. Transfer characteristics of an IGBT-device. Solid lines: Measurement; Symbols: Simulation.

The avalanche breakdown voltage constitutes the maximum voltage a semiconductor device can sustain in the blocking state. It is therefore one of the most important parameters that specifies the safe operating area of the device. For p-n junctions, it is known that the avalanche breakdown voltage increases with increasing temperature due to the lattice scattering rate of the charge carriers. We have measured the static drain-source breakdown voltage for temperatures between room temperature and 400°C. As shown in Figure 9, the breakdown voltage increases linearly with temperature in the whole temperature range investigated. The corresponding simulations were performed using the traditional mobility model of Masetti *et al* [10] as well as the new model developed at the University of Bologna. The parameters used in these simulations were the same as those obtained from the quasi-static simulation of the transfer characteristics (i.e. the simulation data shown in Fig. 10 are based on the re-calibrated models rather than the result of a fit). This example suggests the conclusion that the re-calibrated physical models provide a satisfactory description of MOSFET-like devices.

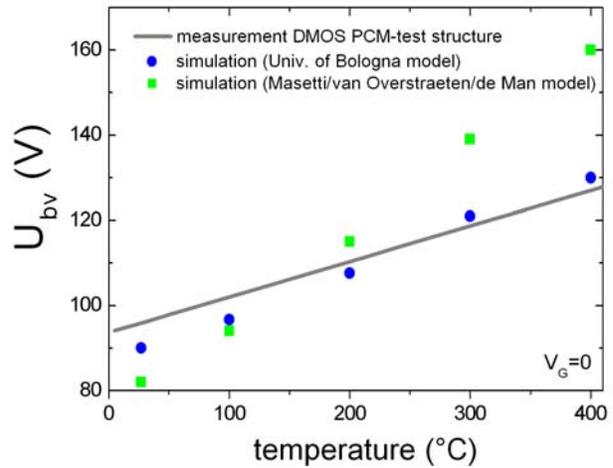

Figure 10. VDMOSFET test structure. Breakdown voltage as a function of temperature. Solid line: measurement; Symbols: simulation.

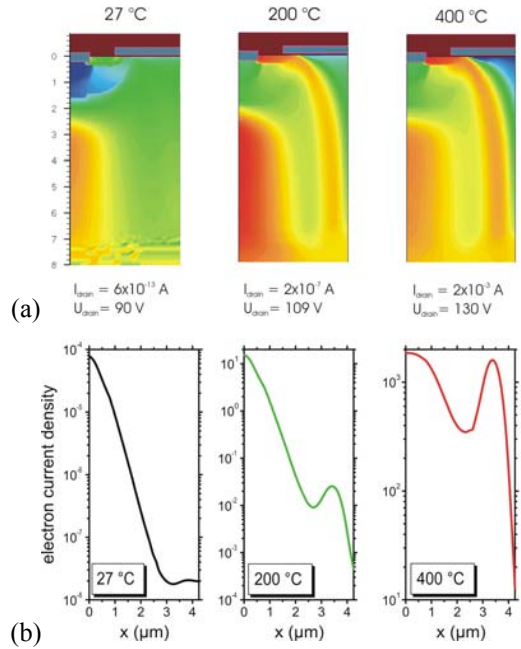

Figure 11. VDMOSFET test structure. a) Simulated electron current density for $U_D = U_{BV}$; b) Simulated electron current density along the x-coordinate line y=3 μm in the upper figure.

More details of the physical process occurring during avalanche breakdown in the interior of these devices are shown in Fig. 11. The distribution of the electron current at the onset of breakdown for temperatures between 27°C and 400°C are displayed in Figure 11a. It demonstrates the uncontrolled shift of the current path from the body-epilayer diode to the parasitic n-p-n transistor indicating the turn on of the latter at about 200°C. Figure 11b shows the radial electron current distribution at a depth of 3 μm. It makes evident that at 400°C a significant part of the avalanche current flows through the parasitic transistor. These facts are closely related to the findings reported by Icaza-Deckelmann *et al.* [11] considering the unclamped inductive switching failure of DMOS-transistors.

For the device simulations we used a version of the commercial simulation tool DESSIS (formerly ISE-TCAD, now Synopsys) [12], in which these extended high temperature models were available. Table 1 summarizes the physical simulation models used in the present work.

TABLE I
PHYSICAL MODELS USED FOR THE DEVICE SIMULATION

| Physical effect | Model used | Reference |
|---|---|---|
| band gap narrowing | Slotboom | [13] |
| bulk mobility | Univ. of Bologna | [2] |
| channel mobility | Lombardi | [14] |
| carrier-carrier scattering | Brooks/Herring | - |
| high field saturation | Canali | [15] |
| recombination | SRH, Auger | - |
| avalanche generation | Univ. of Bologna | [5] |

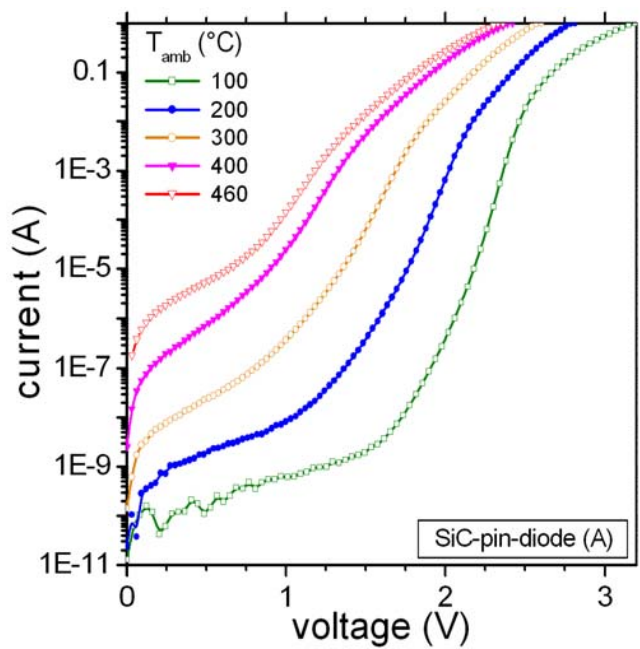

Figure 12. Forward characteristics of a silicon carbide pin-diode for temperatures between 100°C and 460°C.

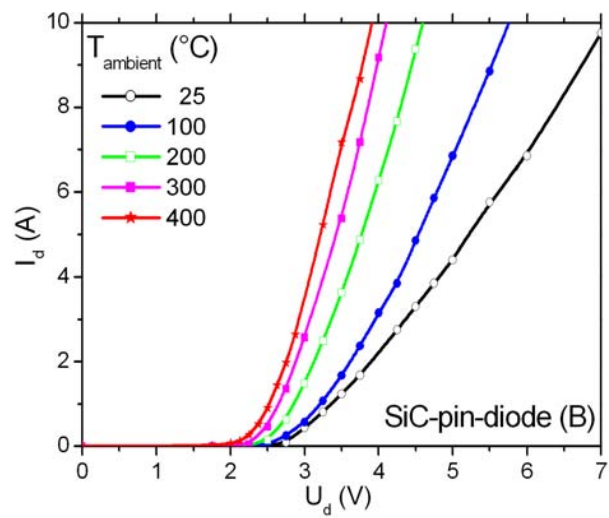

Figure 13. Forward characteristics of another SiC-pin-diode measured with short pulses. The pulse length and period was 100 μs and 50 ms, respectively.

Results for two SiC samples, both delivered by SiCED GmbH (Germany), are shown in figures 12 and 13. Fig. 12 displays the forward characteristics of a SiC pin diode for temperatures between 25° C and 460° C. The diode can be operated up to about 15 A [16], but the maximum current measured was limited by our parameter analyzer to 1A. Pulsed measurements up to 400°C for another sample of the same SiC-pin diode are shown in Fig. 13. More detailed results for the high-temperature behavior of SiC-pin diode may be found in [17].

ACKNOWLEDGMENTS

We would like to thank SiCED GmbH for providing the SiC samples, Infineon for DMOS and IGBT samples and for fruitful collaboration. We also thank H. Schwarzbauer (Siemens AG) for the support with the NTV-method.